\documentclass[prd,reprint,nofootinbib,notitlepage,aps,tightenlines,amsmath,amssymb,showpacs,superscriptaddress]{revtex4-1}

\usepackage[hyperindex,breaklinks]{hyperref}
\usepackage{graphicx,epstopdf,amsmath,amsfonts,amssymb,color,diagbox}

\def\lsim{\mathrel{\rlap{\lower4pt\hbox{\hskip1pt$\sim$}}
    \raise1pt\hbox{$<$}}}                
\def\gsim{\mathrel{\rlap{\lower4pt\hbox{\hskip1pt$\sim$}}
    \raise1pt\hbox{$>$}}}                
\def\be{\begin{equation}}
\def\ee{\end{equation}}
\def\ba{\begin{eqnarray}}
\def\ea{\end{eqnarray}}
\def\ge{\mathrel{\raise.3ex\hbox{$>$\kern-.75em\lower1ex\hbox{$\sim$}}}}
\def\la{\mathrel{\raise.3ex\hbox{$<$\kern-.75em\lower1ex\hbox{$\sim$}}}}

\def\blfootnote{\xdef\@thefnmark{}\@footnotetext}

\def\simgt{\mathrel{\raise.3ex\hbox{$>$\kern-.75em\lower1ex\hbox{$\sim$}}}}
\def\simlt{\mathrel{\raise.3ex\hbox{$<$\kern-.75em\lower1ex\hbox{$\sim$}}}}

\newcommand{\nc}{\newcommand}

\nc{\gone}{\bar g_{\pi NN}^{(1)}}
\nc{\gzero}{\bar g_{\pi NN}^{(0)}}
\nc{\al}{\alpha}
\nc{\ga}{\gamma}
\nc{\de}{\delta}
\nc{\ep}{\epsilon}
\nc{\ze}{\zeta}
\nc{\et}{\eta}
\nc{\ka}{\kappa}
\nc{\rh}{\rho}
\nc{\si}{\sigma}
\nc{\ta}{\tau}
\nc{\up}{\upsilon}
\nc{\ph}{\phi}
\nc{\ch}{\chi}
\nc{\ps}{\psi}
\nc{\om}{\omega}
\nc{\Ga}{\Gamma}
\nc{\De}{\Delta}
\nc{\La}{\Lambda}
\nc{\Si}{\Sigma}
\nc{\Up}{\Upsilon}
\nc{\Ph}{\Phi}
\nc{\Ps}{\Psi}
\nc{\Om}{\Omega}
\nc{\ptl}{\partial}
\nc{\del}{\nabla}
\nc{\ov}{\overline}
\nc{\newcaption}[1]{\centerline{\parbox{15cm}{\caption{#1}}}}
\nc{\us}{U(1)$_S$}

\def\beq{\begin{equation}}
\def\eeq{\end{equation}}
\def\bmat{\begin{displaymath}}
\def\emat{\end{displaymath}}
\def\bear{\begin{eqnarray}}
\def\eear{\end{eqnarray}}
\def\ba{\begin{eqnarray}}
\def\ea{\end{eqnarray}}
\def\bery{\begin{array}}
\def\ery{\end{array}}
\def\bit{\begin{itemize}}
\def\eit{\end{itemize}}
\def\ben{\begin{enumerate}}
\def\een{\end{enumerate}}
\def\btab{\begin{tabular}}
\def\etab{\end{tabular}}
\def\btbl{\begin{table}}
\def\etbl{\end{table}}
\def\bfig{\begin{figure}[htb]}
\def\efig{\end{figure}}
\def\bpic{\begin{picture}}
\def\epic{\end{picture}}

\def\ga{\mathrel{\raise.3ex\hbox{$>$\kern-.75em\lower1ex\hbox{$\sim$}}}}
\def\la{\mathrel{\raise.3ex\hbox{$<$\kern-.75em\lower1ex\hbox{$\sim$}}}}
\def\gappeq{\mathrel{\rlap {\raise.5ex\hbox{$>$}}
{\lower.5ex\hbox{$\sim$}}}}
\def\lappeq{\mathrel{\rlap{\raise.5ex\hbox{$<$}}
{\lower.5ex\hbox{$\sim$}}}}

\def\gyr{{\rm \, G\kern-0.125em yr}}
\def\mev{{\rm \, Me\kern-0.125em V}}
\def\gev{{\rm \, Ge\kern-0.125em V}}
\def\tev{{\rm \, Te\kern-0.125em V}}

\renewcommand{\bar}{\overline}

\newcommand{\rst}[1]{\raise+.6ex\hbox{#1}}

\begin{document}


\title{The Leptonic Higgs Portal}

\author{Brian Batell}
\affiliation{Pittsburgh Particle Physics, Astrophysics, and Cosmology Center,
Department of Physics and Astronomy, University of Pittsburgh, PA 15260, USA}

\author{Nicholas Lange}
\affiliation{Department of Physics and Astronomy, University of Victoria, 
Victoria, BC V8P 5C2, Canada}

\author{David McKeen}
\affiliation{Department of Physics, University of Washington, Seattle, WA 98195, USA}

\author{Maxim Pospelov}
\affiliation{Department of Physics and Astronomy, University of Victoria, 
Victoria, BC V8P 5C2, Canada}
\affiliation{Perimeter Institute for Theoretical Physics, Waterloo, ON N2J 2W9, 
Canada}

\author{Adam Ritz}
\affiliation{Department of Physics and Astronomy, University of Victoria, 
Victoria, BC V8P 5C2, Canada}

\date{June 2016}


\begin{abstract} 
An extended Higgs sector may allow for new scalar particles well below the weak scale. 
In this work, we present a detailed study of a light scalar $S$ with enhanced coupling to leptons, which could be responsible for the 
existing discrepancy between experimental and theoretical determinations of the muon anomalous magnetic moment. We present an ultraviolet completion 
of this model in terms of the lepton-specific two-Higgs doublet model and an additional scalar singlet. We then analyze a plethora of experimental constraints on the universal low energy model, and this  UV completion, along with the
sensitivity reach at future experiments. The most relevant constraints originate from muon and kaon decays, electron beam dump experiments, electroweak precision observables, rare $B_d$ and $B_s$ decays and Higgs branching fractions. 
The properties of the leptonic Higgs portal imply an enhanced couplings to heavy leptons, and we identify the 
most promising search mode for the high-luminosity electron-positron colliders as $e^+{+}e^-\to\tau^+{+}\tau^-{+}S \to \tau^+{+}\tau^-{+}\ell{+}\bar \ell$, 
where $\ell =e,\mu$.
Future analyses of existing data from BaBar and Belle, and from the upcoming Belle II experiment, 
 will enable tests of this model as a putative solution to the muon $g-2$ problem for $m_S < 3.5$ GeV.
\end{abstract}
\maketitle


\section{Introduction}
\label{sec:intro}

The LHC discovery of a new particle of mass $\sim$125~GeV, with properties consistent with those of the Standard Model Higgs boson~\cite{Aad:2012tfa,*Chatrchyan:2012ufa}, provides compelling evidence for the picture of the electroweak symmetry, and its spontaneous breakdown, encapsulated in the Standard Model (SM). It remains an important question to understand whether the entire Higgs sector is minimal, as in the SM, or contains additional states as would be required by supersymmetry, or may be motivated by other scenarios including, for example, models of dark matter. 

While the existence of new physics at the TeV scale is still a distinct possibility (see e.g.~\cite{750GeV}), in recent years, independent empirical motivations related to dark matter and neutrino masses have pointed to the possibility of a hidden sector, weakly coupled to the SM~\cite{Essig:2013lka}. 
The mass scales in the hidden sector can be considered free 
parameters, and therefore particles much lighter than the electroweak or TeV scales are plausible. On general effective field theory grounds, the leading interactions with a neutral light hidden sector would be through the relevant and marginal interactions involving SM gauge singlets, which have been dubbed ``portals"~\cite{Patt:2006fw} and are the subject of considerable theoretical and experimental study.

In several cases, hypothetical light particles may help to explain certain experimental anomalies and deviations from the SM. It has been appreciated that a rather minimal extension of the SM via an additional vector particle $V$ (often termed the ``dark photon") that kinetically mixes with the photon through the interaction $(\epsilon/2)V^{\mu\nu}F_{\mu\nu}$, where $V^{\mu\nu}$ and $F^{\mu\nu}$ are the $V$ and photon field strengths respectively, can generate an appreciable shift of the muon anomalous magnetic moment~\cite{Pospelov:2008zw},
\be
\Delta a_\mu \simeq \frac{ \alpha \epsilon^2}{2 \pi}~~{\rm ~ when }~ m_V\ll m_\mu.
\ee
For $\epsilon \sim 10^{-3}$, such a model offers a correction 
on the order of the existing discrepancy in $a_\mu$, with the 
right sign to alleviate the tension between theory and experiment~\cite{Bennett:2006fi}. A subsequent 
painstaking search for light dark photons in both old data and in dedicated new experiments has resulted in upper limits on $\epsilon$ 
that now render the {\em minimal} dark photon model unable to explain the existing discrepancy. (The last remaining portion of the parameter space able to account for the discrepancy was excluded by the NA48/2 experiment~\cite{Batley:2015lha}.) However, modifications of the minimal vector
portal model, for example dark photons decaying to other dark sector states, and gauge groups based on $L_\mu - L_\tau$, are still able to 
shift $a_\mu$ by $3\times 10^{-9}$ (the scale of the experimental discrepancy), and be consistent with all other constraints (see, e.g., \cite{Gninenko:2001hx,Batell:2009di,Batell:2014mga,Altmannshofer:2014pba}). 

\begin{figure*}[t]
\centerline{\includegraphics[width=.48\textwidth]{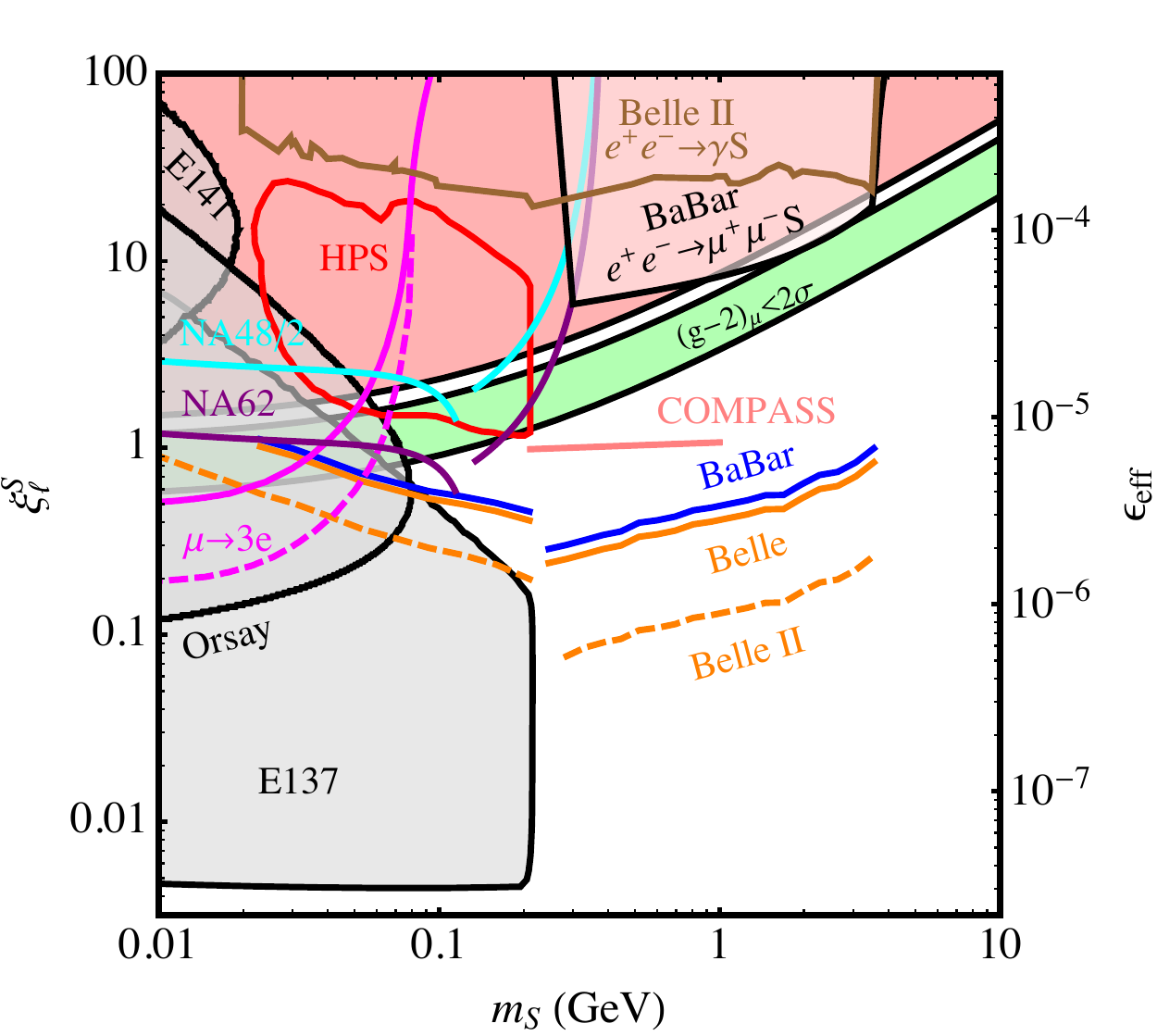}\hspace*{0.5cm}\includegraphics[width=.48\textwidth]{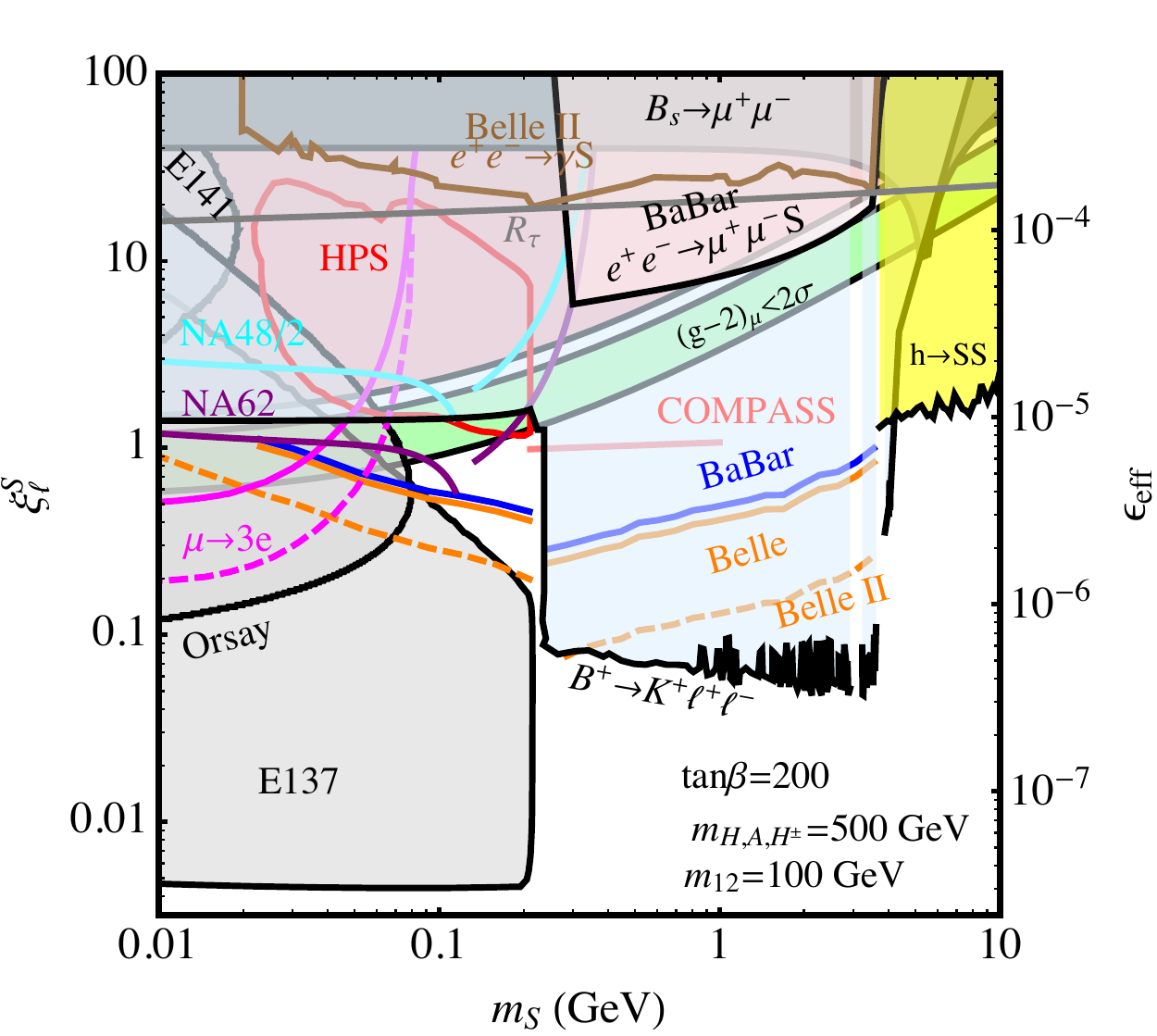}}
\caption{\footnotesize {\it Left panel:} Constraints on the coupling to leptons (in terms of both $\xi^S_\ell=g_\ell(v/m_\ell)$ and $\epsilon_{\rm eff}=g_e/e$) as a function of the scalar mass, based purely on the effective theory in Eq.~(\ref{noUV}). The region where $(g-2)_\mu$ is discrepant at $5\sigma$ is shaded in red, while the green shaded band shows where the current discrepancy is brought below $2\sigma$. We show constraints from the beam dumps E137, Orsay, and E141. The projected sensitivities from $\mu\to3e$, NA48/2, NA62, HPS, analyses of existing data from COMPASS and $B$-factories, as well as a projected sensitivity at BELLE II are also shown. (See Section~3 for details.) {\it Right panel:} Constraints on the L2HDM+$\varphi$ UV completion of the effective theory in Eq.~(\ref{noUV}), as described in Sec.~\ref{sec:L2HDMS}. Model independent results are as in the left panel. In addition, for this particular UV completion, there are constraints on the model from searches for $h\to SS\to 2\mu2\tau$, $B\to K^{(\ast)}\ell^+\ell^-$, and $B_s\to \mu^+\mu^-$. We have set $\tan\beta=200$, $m_{H}=m_{H^\pm}=500~\rm GeV$, and  $m_{12}=1~\rm TeV$. (See Section~4 for details.)}
\label{Cons}
\end{figure*}

In this paper, we concentrate on light scalars coupled to leptons as a prospective solution to the muon $g-2$ anomaly.
The relevant observation was originally made by Kinoshita and Marciano \cite{Kinoshita:1990aj}: a SM-like Higgs boson with a very light mass,
$m_h \ll m_\mu$ (excluded by now via numerous experiments culminating in the 
discovery of the Higgs at the LHC), gives the following positive shift of the muon anomalous magnetic moment,
\be
\Delta a_\mu =  \frac{3}{16 \pi^2} \times \left( \frac{m_\mu}{v} \right)^2  \simeq 3.5 \times 10^{-9},
\ee
which is very close to the existing discrepancy. In this expression, $v=246~\rm GeV$ is related to the vacuum expectation value of the Higgs doublet, $H$, via $\langle H\rangle=v/\sqrt2$. The lesson of this observation is that if a new light scalar particle
couples to leptons with a coupling strength on the order of the SM lepton Yukawa couplings, which in the case of the muon is
$m_\mu/v \simeq 4\times 10^{-4}$, the muon $g-2$ problem can be solved. 
Thus we are motivated to study the effective Lagrangian of an elementary scalar $S$, 
\be
{\cal L}_{\rm eff} = \frac{1}{2}(\partial_\mu S)^2 - \frac{1}{2}m_S^2 S^2 +\sum_{l = e,\mu,\tau} g_\ell S \bar\ell\ell,
\label{noUV}
\ee
with $g_l \sim m_l/v$ as a promising phenomenological model. Given that $S$ is not the SM Higgs boson, the interaction terms in 
(\ref{noUV}) may appear to contradict SM gauge invariance. Thus, at minimum, Eq.~(\ref{noUV}) requires 
an appropriate UV completion, generically in the form of new particles at the electroweak (EW) scale charged under the SM gauge group. 
On the other hand, {\em  if} a UV-complete model is found that represents a consistent generalization of (\ref{noUV}), 
the light scalar solution to the muon $g-2$ problem deserves additional attention. 
 Another impetus for studying very light beyond-the-SM (BSM) scalars comes from the 
existing discrepancy of the muon- and electron-extracted charge radius of the proton~\cite{TuckerSmith:2010ra}.

This paper presents a detailed study of  light scalars with enhanced coupling to leptons, and
provides a viable UV-completion of Eq. (\ref{noUV}) through what we dub the `leptonic Higgs portal'. We also analyze a variety of phenomenological consequences of the model. 
The phenomenology of a light scalar coupled to leptons resembles in many ways the phenomenology of the dark photon, but with the distinct feature that the couplings to individual 
flavors are non-universal and proportional to the mass. As a result, at any given energy the production of such a scalar is most efficient using the heaviest 
kinematically accessible lepton. We identify the most important search modes for the scalar that could decisively explore its low mass regime. Our main conclusion is that an elementary scalar with coupling to leptons $\ell$ scaling as $m_\ell$ can be very efficiently probed, and in particular
the whole mass range consistent with a solution of the muon $g-2$ discrepancy can be accessed through an analysis of existing data and in upcoming experiments.

 Our full UV-complete model is based on the lepton-specific two Higgs doublet model with an additional light scalar singlet. The mixing of the singlet with components of the electroweak doublets results in the effective Lagrangian of Eq.~(\ref{noUV}). The model also induces additional observables, and thus constraints, due to the fact that $S$ receives small but nonvanishing couplings to the 
SM quarks and gauge bosons. We note that the UV completion presented in this work is not unique. For an alternative UV completion of the same model utilizing vector-like fermions at the weak scale, see Ref.~\cite{Chen:2015vqy}. 
While many aspects of the low-energy phenomenology based on the 
effective Lagrangian (\ref{noUV}) are similar in both approaches, the 
UV-dependent effects are markedly different (especially for  flavor-changing observables). 

This paper is organized as follows. In the next section we discuss light scalars coupled to leptons and 
a possible UV completion of such models via the leptonic Higgs portal. 
In Sec.~\ref{sec:model_indep} we analyze the constraints and sensitivity levels to light scalars coupled to leptons that are universal, and independent of the 
UV completion (resulting from muon decays, leptonic kaon decays, electron beam dumps and high-intensity $e^+e^-$ colliders); the results are shown in the left panel of Fig.~\ref{Cons}. In Sec.~\ref{sec:model_dep} we  analyze the constraints and sensitivities that are tied to the specific UV-completion involving the leptonic Higgs portal. These include rare $B$ and Higgs decays; the results are shown in the right panel of Fig.~\ref{Cons}.
We present some additional discussion and reach our conclusions in Sec.~\ref{sec:concl}.

\section{Leptonic Higgs portal} 
\label{sec:L2HDMS}
In this section, we discuss a concrete UV-completion of the low-energy Lagrangian in Eq.~(\ref{noUV}). A simple starting point to couple a singlet field $\varphi $ to the SM is through the Higgs portal,
\be
{\cal L}_{\rm int} =(A \varphi +\lambda \varphi ^2)H^\dagger H,
\label{Hportal}
\ee 
where $H$ is the SM Higgs doublet and $A,\lambda$ are coupling constants.
The trilinear term induces mixing between the singlet and the ordinary Higgs boson $h$ after electroweak symmetry breaking, where $H^0 = (v+h)/\sqrt{2}$. The mixing angle is given by
\be
\theta = \frac{Av}{m_h^2 - m_\varphi ^2},
\label{mixing}
\ee
and after field diagonalization the coupling of the light scalar $S$ (mostly comprised of the singlet $\varphi$) to SM fermions is simply given by their SM Yukawa coupling times this mixing angle. 
Low mass singlets are constrained by $B$ and $K$ meson decays (see, e.g. a collection of theoretical and experimental studies in Refs.~\cite{Bird:2004ts,O'Connell:2006wi,Pospelov:2007mp,Batell:2009jf,Bezrukov:2009yw,Aaij:2013lla,Aaij:2015tna,Krnjaic:2015mbs}), and for $m_S < 4~\rm GeV$
the mixing angle is limited to $|\theta| < 10^{-3}$. Significant further advances in sensitivity to $\theta$ are possible 
with the planned SHiP experiment \cite{Alekhin:2015byh}. Therefore, there is no room to accommodate $\theta \sim O(1)$, and 
consequently no large correction to the muon $g-2$ is allowed within this simple model. 

To circumvent this obstacle, we modify the SM by not only adding a singlet but also by introducing a second Higgs doublet that  mixes with the 
singlet. In particular, we are interested in the so-called `lepton-specific' representation of a generic two Higgs doublet model (L2HDM)~\cite{Su:2009fz,Marshall:2010qi,Branco:2011iw,Chun:2015hsa,Chun:2016hzs}. 
Calling the two doublets with SM Higgs charge assignments $\Phi_1$ and $\Phi_2$, 
we assume that  $\Phi_1$ couples exclusively to leptons, while $\Phi_2$ couples to quarks. 
Moreover, we assume that all physical compenents of $\Phi_{1,2}$ are at the weak scale or above.
Taking $\langle \Phi_2\rangle /\langle \Phi_1 \rangle \equiv \tan \beta $ very large, as well as arranging for the physical bosons of $\Phi_1$ to be heavier than those of $\Phi_2 $, 
we arrive at an ``almost SM-like" limit, but with the set of heavier Higgses that couple to leptons possessing couplings enhanced by $\tan\beta$. 
The mixing term $A_{12}(\Phi_1^\dagger \Phi_2 +\Phi_2^\dagger \Phi_1)\varphi $ will then efficiently mix $\varphi $ with $\Phi_1$, resulting in the light scalar $S$ coupling to leptons with strength
\be
g_\ell =  \frac{m_\ell}{v} \times \tan \beta  \times \theta_\ell,
\label{mixing_l}
\ee
where $\theta_\ell$ is the mixing between $S$ and $\Phi_1$. It is then clear that the desirable outcome of $g_\ell \sim  m_\ell/v$ can be achieved in the 
regime $\tan\beta\gg 1$, $\theta_\ell \ll 1$, and $\tan \beta  \times \theta_\ell \sim O(1)$. 

We now elaborate on this simple idea and present details of the model. The scalar potential we consider 
is given by 
\be
V(\Phi_1, \Phi_2,\varphi ) = V_{2\rm HDM} + V_\varphi + V_{\rm portal}.
\label{sc_potential}
\ee
$V_{2\rm HDM}$ is the main part of the potential that determines the pattern of electroweak symmetry breaking. 
Its CP-conserving version is given by the familiar expression,
\begin{align}
V_{2\rm HDM}&=m_{11}^2\Phi_1^\dagger\Phi_1+m_{22}^2\Phi_2^\dagger\Phi_2-m_{12}^2\left(\Phi_1^\dagger\Phi_2+\Phi_2^\dagger\Phi_1\right) \nonumber\\
 &\hspace*{-0.7cm}+\frac{\lambda_1}{2}\left(\Phi_1^\dagger\Phi_1\right)^2+\frac{\lambda_2}{2}\left(\Phi_2^\dagger\Phi_2\right)^2 +\lambda_3\left(\Phi_1^\dagger\Phi_1\right)\left(\Phi_2^\dagger\Phi_2\right) \nonumber\\
  &\hspace*{-0.7cm}+\lambda_4\left(\Phi_1^\dagger\Phi_2\right)\left(\Phi_2^\dagger\Phi_1\right)+\frac{\lambda_5}{2}\left[\left(\Phi_1^\dagger\Phi_2\right)^2+\left(\Phi_2^\dagger\Phi_1\right)^2\right].
\label{eq:V2HDM}
\end{align}
The singlet potential in (\ref{sc_potential}) is a generic polynomial with positive $\varphi^4$ term, 
\be
V_S=B\varphi +\frac{1}{2}m_0^2\varphi ^2+\frac{A_\varphi }{2}\varphi ^3+\frac{\lambda_\varphi }{4}\varphi ^4.
\label{eq:VS}
\ee
In the portal part of the potential we are most interested in the trilinear terms, 
\begin{align}
\nonumber
V_{\rm portal} 
=\left[A_{11}\Phi_1^\dagger\Phi_1+A_{22}\Phi_2^\dagger\Phi_2 \right.~~~~~~~~~
\\\left.+A_{12}\left(\Phi_1^\dagger\Phi_2+\Phi_2^\dagger\Phi_1\right)\right]\varphi.
\end{align}
Generically, the $A_{11}$ portal term leads to a $1/\tan\beta$ suppressed mixing between $\varphi$ and the electroweak scalars, 
while, for $\tan\beta\gg 1$, the $A_{22}$ portal coupling is strongly 
constrained by existing limits on the $A\varphi H^\dagger H$ operator. On the other hand, the $A_{12}$ portal is less constrained and leads to efficient mixing with $\varphi$. In what follows we will ignore $A_{11,22}$.

The spectrum of the theory at the electroweak scale is dominated by $V_{2\rm HDM}$, while $V_\varphi$ and $V_{\rm portal}$ can be regarded as small perturbations. 
In determining the spectrum at the weak scale, we decompose the doublets assuming each obtains a vacuum expectation value, $\langle \Phi_a\rangle\equiv v_a$,
\begin{align}
\Phi_a&\supset v_a+\rho_a
\end{align}
for $a=$1, 2 with $\rho_a$ a real scalar field (we work in unitary gauge and ignore charged components of the doublets for now). The ratio of the VEVs is $v_2/v_1=\tan\beta$ with $v_1^2+v_2^2\equiv v^2=(246~{\rm GeV})^2$. Furthermore, through a $\varphi$ field redefinition, the coefficient $B$ in Eq. (\ref{eq:VS}) can be chosen so that $\varphi$ does not obtain a VEV.

The elements of the mass matrix of the neutral CP-even scalars in the basis $\left(\rho_1,\,\rho_2,\,\varphi \right)$ are
\begin{align}
M^2_{11}&=m_{12}^2\tan\beta+\lambda_1v^2\cos^2\beta,
\label{eq:2HDMmass1}
\\
M^2_{22}&=m_{12}^2\cot\beta+\lambda_2v^2\sin^2\beta,
\label{eq:2HDMmass2}
\\
M^2_{12}&=-m_{12}^2+\lambda_{345}v^2\cos\beta\sin\beta,
\label{eq:2HDMmass3}
\\
M^2_{13}&=v A_{12}\sin\beta,~M^2_{23}=vA_{12}\cos\beta,~M^2_{33}=m_{0}^2.
\end{align}
with $\lambda_{345}\equiv\lambda_3+\lambda_4+\lambda_5$. In the limit that $A_{12}\ll v,\, m_{12}$, we can rotate to the mass basis perturbatively,
\begin{align}
\left(
\begin{array}{c}
\rho_1 \\
\rho_2 \\
 \varphi 
\end{array}
\right)\simeq\left( 
\begin{array}{ccc}
-\sin\alpha & \cos\alpha & \delta_{13} \\
\cos\alpha & \sin\alpha & \delta_{23} \\
\delta_{31} & \delta_{32} & 1
\end{array}
\right)
\left( 
\begin{array}{c}
h \\
H \\
S
\end{array}
\right),
\end{align}
with small mixing angles $\delta_{ij}$, and $\alpha$ satisfying
\begin{align}
\tan 2\alpha=\frac{2M_{12}^2}{M_{11}^2-M_{22}^2}.
\label{eq:alpha}
\end{align}
The masses of the physical states $h$ and $H$ are
\begin{align}
m_{h,H}^2&=\frac{1}{2}\bigg[M_{11}^2+M_{22}^2
\label{eq:doubletmasses}
\\
&\quad\quad\mp\sqrt{\left(M_{11}^2-M_{22}^2\right)^2+4\left(M_{12}^2\right)^2}\bigg],
\nonumber
\end{align}
while the mass of $S$ is
\begin{align}
m_S^2&\simeq m_0^2+\delta_{13}M^2_{13}+\delta_{23}M^2_{23}.
\label{eq:Smass}
\end{align}
We will see that $S$ can be rendered light while coupling dominantly to leptons (when $\tan\beta$ is large) below, putting off questions of fine-tuning for the time being.

In the L2HDM, the Yukawa interactions of $\Phi_1$ and $\Phi_2$ with fermions are given by
\begin{align}
-{\cal L}_Y=\bar L Y_e\Phi_1 e_R+\bar Q Y_d\Phi_2 d_R+\bar Q Y_u\tilde\Phi_2 u_R+{\rm h.c.},
\label{eq:doubletyukawas}
\end{align}
suppressing generational indices and using first generation notation. The Yukawa content of this model is exactly the same as in the SM, ensuring a pattern of minimal flavor violation (MFV). In particular, there are no flavor-changing neutral currents (FCNCs) mediated by either of the Higgs fields at tree level.
The only difference with the SM is through the appearance of the vacuum angle $\beta$ in the mass-Yukawa coupling relation, 
\be
m_e = \cos\beta \times \frac{Y_e v}{\sqrt{2}},~~ m_{u(d)} = \sin\beta \times \frac{Y_{u(d)} v}{\sqrt{2}}.
\ee
In the large $\tan\beta$ regime, the size of the Yukawa couplings in the quark sector is almost the same as in the 
SM, but in the lepton sector all Yukawa couplings are enhanced by $\tan\beta$. 

Upon diagonalization of the Higgs mass matrix, the Yukawa interactions of the physical states are
\begin{align}
\label{xi_psi}
-{\cal L}_Y&\supset\sum_{\substack{\phi=S,\,h,\,H\\ \psi=\ell,\,q}}\xi^\phi_{\psi}\frac{m_\psi}{v}\phi\bar\psi\psi
\end{align}
where $\ell$ labels each generation of lepton fields and $q$ those of the quarks. The couplings to the weak gauge bosons can be found by expanding the kinetic terms of the doublets in the Lagrangian or by expanding $v^2$ about the vacuum:
\begin{align}
\label{xi_V}
{\cal L}&\supset\sum_{\phi=S,\,h,\,H}\xi^\phi_{V}\frac{\phi}{v}
\left(2m_W^2W_\mu^+W^{\mu -}+m_Z^2Z_\mu Z^\mu\right).
\end{align}
Defined this way, $\xi^\phi_{\psi,V}=1$ is a coupling of SM Higgs strength.
In Table~\ref{tab:couplings}, we show these couplings in terms of the angles $\alpha$ and $\beta$, and in Table~\ref{tab:couplingsapprox} provide approximate values in the regime of interest.
\begin{table}
\begin{center}
\begin{tabular}{|c|c|c|c|} \hline
\diagbox{$\psi$}{$\phi$} & $S$ & $h$ & $H$ \\ \hline
$\ell$ & $\delta_{13}/c_\beta$ & $-s_\alpha/c_\beta$ & $c_\alpha/c_\beta$ \\
$q$ & $\delta_{23}/s_\beta$ & $c_\alpha/s_\beta$ & $s_\alpha/s_\beta$ \\
$W$, $Z$ & $\delta_{13}c_\beta+\delta_{23}s_\beta$ & $\sin\left(\beta-\alpha\right)$ & $\cos\left(\beta-\alpha\right)$ \\ \hline
\end{tabular}
\caption{Values of $\xi^\phi_{\psi}$ for $\phi=S$, $h$, $H$, $\psi=\ell$, $q$, $W$, $Z$ in the L2HDM+$\varphi$.
\label{tab:couplings}}
\end{center}
\end{table}

\begin{table}
\begin{center}
\begin{tabular}{|c|c|c|c|} \hline
\diagbox{$\psi$}{$\phi$} & $S$ &  $h$ & $H$  \\ \hline
$\ell$ & $-\left(vA_{12}/m_{H}^2\right)\tan\beta$ & $\xi^h_{\ell}$ & $\pm\tan\beta$  \\
$q$ & $-\left(vA_{12}/m_h^2\right) x \cot\beta$ & $1$ & $\mp\xi^h_{\ell}\cot\beta$  \\
$W$, $Z$ & $-\left(vA_{12}/m_h^2\right)\left(r+x\right)\cot\beta$ & $1$ & $\pm\left(1-\xi^h_{\ell}\right)\cot\beta$ \\ \hline
\end{tabular}
\caption{Approximate values of $\xi^\phi_{\psi}$ when $\tan\beta\gg 1$ for $\phi=S$, $h$, $H$, $\psi=\ell$, $q$, $W$, $Z$, with $\alpha$ chosen so that $\xi^h_{\ell}\simeq 1$, $r\equiv m_h^2/m_{H}^2$ and $x\equiv1+\xi^h_{\ell}\left|1-r\right|$ in the L2HDM+$\varphi$ for $m_h<m_H$ ($m_h>m_H$). 
\label{tab:couplingsapprox} }
\end{center}
\end{table}

We assume that $h$ has SM-like couplings to the gauge bosons and quarks, which means that $\cos\left(\beta-\alpha\right)\simeq 0$ and $\cos\alpha\simeq\sin\beta$. Furthermore, if $\tan\beta\gg 1$, then $H$ and $S$ will couple much more strongly to leptons than to quarks. 
This can be accomplished by choosing $\alpha\simeq 0$ (and negative) and $\beta\simeq\pi/2$. 
In this case, we can make $h$ arbitrarily SM-like, consistent with the observations of the ATLAS and CMS experiments, 
while allowing $m_H$ and $\tan\beta$ to vary (again ignoring questions of fine tuning for now).

Given this pattern of masses and couplings, we can find the singlet mixing angles,
\begin{align}
\delta_{13}&\simeq-\frac{vA_{12}}{m_{H}^2},~\delta_{23}\simeq-\frac{vA_{12}}{m_h^2}\left[1+\xi^h_{\ell}\left(1-\frac{m_h^2}{m_H^2}\right)\right]\cot\beta,
\label{eq:deltas}
\end{align}
or
\begin{align}
\xi^S_\ell&\simeq-\frac{vA_{12}}{m_{H}^2}\tan\beta,
\\
\xi^S_q&\simeq-\frac{vA_{12}}{m_h^2}\left[1+\xi^h_{\ell}\left(1-\frac{m_h^2}{m_H^2}\right)\right]\cot\beta.
\end{align}
Recall that the Yukawa couplings of $S$ are $g_{\ell,q}=\xi^S_{\ell,q} m_{\ell,q}/v$.

We can re-express the mass shift of the lightest scalar from Eq.~(\ref{eq:Smass}) due to electroweak symmetry breaking in terms of  more physical parameters,
\begin{align}
m_S^2&\simeq m_0^2-\left(\frac{m_H \xi^S_\ell}{\tan\beta}\right)^2.
\end{align}
The cancellation between $\delta m_S^2$ and $m_0^2$ to obtain a GeV-scale value of $m_S$ represents a (mild) 
fine-tuning in this theory. We have checked that the hierarchy of the mass scales, $m_S \ll m_{h,H}$ is indeed possible without 
inducing an instability of the corresponding minimum in the scalar potential.

\section{Universal constraints on the (leptonic) light scalar} 
\label{sec:model_indep}
We subdivide all the possible constraints on the light scalar $S$ into two groups. The first, {\it model independent}, group relies exclusively on the coupling to leptons in Eq.~(\ref{noUV}), and comes mostly from low and medium energy processes, and does not use any of the additional particles brought in by the UV completion. We present the second, {\it model dependent}, group of constraints in the next Section.

Although we introduced the notation $g_\ell=\xi^S_\ell m_\ell/v$ in describing a particular UV completion in Sec.~\ref{sec:L2HDMS}, we will make use of this parameterization and present results in this Section in terms of $\xi^S_\ell$, i.e. normalizing $g_\ell$ on the SM Higgs Yukawa coupling.
\subsection{Lifetimes and decay modes of $S$}

We will concentrate on the masses in the range $1~\rm MeV$ to a few GeV for $m_S$. 
(A region from $\sim$ 200 keV to $2m_e\simeq 1$ MeV may represent an interesting 
blind spot \cite{Izaguirre:2014cza,Liu:2016qwd}, but is not treated in this paper.)
In this mass range, the dominant decay modes of $S$ are to leptons,  
with partial width given by 
\be
\Gamma_{S\to \ell\bar \ell} = g_\ell^2 \times \frac{m_S}{8 \pi} \left(1-\frac{4 m_\ell^2}{m_S^2}\right)^{3/2}.
\label{Ga_S} 
\ee
Depending on the coupling strength and the boost of the $S$ particle produced, the decay length of $S$ can be macroscopic, or rather prompt. 
For example, for $m_S = 1$~GeV, the proper decay length is 
\be
c\tau(m_S= 1\,{\rm GeV}) \simeq 3 \times 10^{-6} {\rm cm} \times \left( \frac{1}{\xi^S_\ell} \right)^2,
\ee
and the decay is prompt. 

\begin{figure}
\centerline{\includegraphics[width=0.5\textwidth]{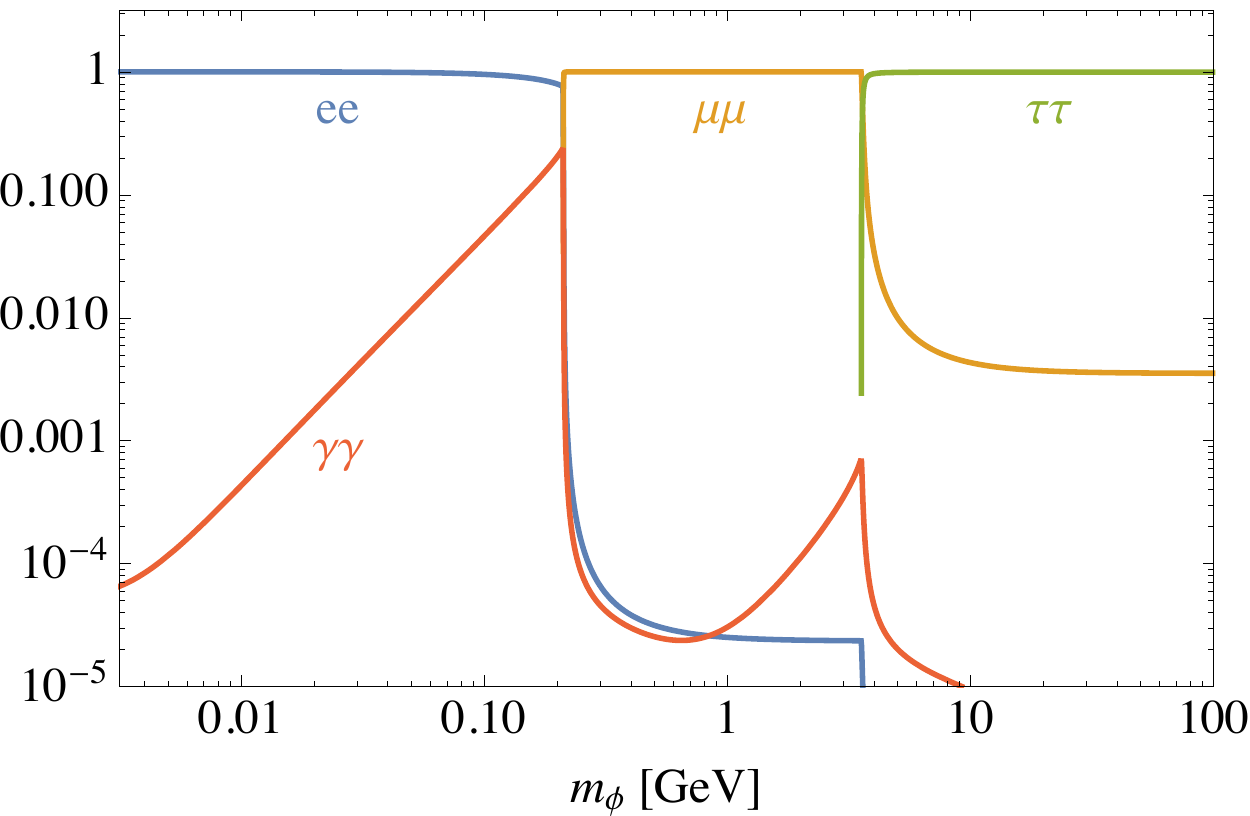}}
\caption{Branching ratios for $S\to\gamma\gamma$, $e^+e^-$, $\mu^+\mu^-$, $\tau^+\tau^-$ as a function of $m_S$.}
\label{branchings}
\end{figure}

The $\gamma\gamma$ decay fraction may become noticeable (up to $\sim 20\%$ just below $m_S = 2m_\mu$) due to the loop-induced 
coupling to photons. In our model, the scaling $g_\ell \propto m_\ell$ allows for unambiguous determinations of the corresponding branching ratios. 
We plot the branching ratios of $S$ as a function of its mass in Fig.~\ref{branchings} noting that the decay is always dominated by the heaviest kinematically allowed lepton pair.

\subsection{Muon anomalous magnetic moment}
A loop of light scalars contributes to the anomalous magnetic moments of fermions. 
A straightforward calculation gives 
\be
a_\ell = \frac{g_\ell^2}{8\pi^2}\int_0^1\frac{(1-z)^2(1+z)}{(1-z)^2 + z(m_S/m_\ell)^2},
\label{g-2}
\ee
which, in the limits of a very light and a very heavy scalar, reduces to $3g_\ell^2/(16\pi^2)$ and $g_\ell^2/(4\pi^2)\times (m_\ell^2/m_S^2) \log(m_S/m_\ell)$ respectively. 
Equation~(\ref{g-2}) and the $g_\ell \propto m_\ell$ dependence lead to $a_\ell$ scaling as the second (fourth) power of lepton mass in the limit of a light (heavy) scalar. 
The tau lepton $g-2$ receives the largest contribution from scalar exchange, but is not measured to the required precision 
(and, in fact, the $a_\tau$ sign is not experimentally determined).
The strongest constraints come from $g-2$ of the muon, and if the the current discrepancy, which we take to be $\left(26.1\pm8.0\right)\times10^{-10}$~\cite{Hagiwara:2011af}, is interpreted as new physics, it suggests a 
non-zero range for $\xi^S_\mu$ shown in Fig.~\ref{Cons}. Notice that, in contrast to the dark photon case, the highly precise measurements of electron $g-2$ do not provide competitive sensitivity. For the rest of the paper, we will treat the suggested muon $g-2$ band as a target of opportunity, and investigate 
other observables that could provide complementary sensitivity to $g_\mu$ in this range. 

To facilitate comparison with the dark photon case, we show results in Fig.~\ref{Cons} (left panel) in terms of both $\xi^S_\ell=g_e(v/m_e)$ and $\epsilon_{\rm eff}\equiv g_e/e$, where $-e$ is the charge of the electron, which is the coupling strength to the electron of a dark photon with kinetic mixing angle $\epsilon_{\rm eff}$. Expressed in terms of $\epsilon_{\rm eff}$, regions determined by the coupling to the electron are in roughly the same place as those in the dark photon case (modulo small differences due to scalar vs. vector properties), while those determined by couplings to $\mu$ and $\tau$ move to smaller values of $\epsilon_{\rm eff}$ by factors of $\sim m_{\mu,\tau}/m_e$.

Note that in our UV-completion via the leptonic Higgs portal, there are additional contributions to $a_\mu$ from the heavy neutral and charged Higgs states. These contributions are subdominant to that of $S$, unless some of the neutral scalars are light, below the mass of the weak bosons. In this work, we will assume the heavy Higgs bosons are much heavier than this, so that the dominant contribution to $a_\mu$  comes from $S$, but see {\it e.g.} 
Refs.~\cite{Abe:2015oca,Chun:2015hsa,Chun:2016hzs} for a recent study exploring this region of parameter space in the lepton-specific 2HDM.

\subsection{ Beam-dump and fixed target constraints}

The coupling of the scalar $S$ to electrons is considerably smaller than to muons, $g_e/g_\mu = (m_e/m_\mu)  \simeq 0.005$.
Consequently, low mass scalars with $m_S < 2m_\mu$  can have displaced decays, or even travel a 
macroscopic distance before decaying. Fig.~\ref{Cons} shows constraints from older beam dump experiments, such as E137 and E141. 
In both cases, the scalars $S$ are produced in an underlying bremsstrahlung-like process, $e + {\rm Nucleus} \to e + S + {\rm Nucleus}$. Notice that these experiments 
firmly rule out scalars with masses below 30 MeV as candidates for the solution of muon $g-2$ discrepancy. Consequently, for the rest of the constraints, 
we will concentrate on $m_S > 10$ MeV. It is also important to note the modification of the 
shape of the excluded region compared to the case of dark photons, universally coupled to all leptons. In the scalar model above 
$m_S=210$ MeV, there is no sensitivity in the beam dump experiments due to abrupt shortening of the lifetime of $S$ by the 
muon pair decay channel. 

The JLab experiment HPS~\cite{Battaglieri:2014hga} utilizes a fixed target, scattering electrons on tungsten, producing scalars through their couplings to electrons. It has the capability to detect displaced 
decays within a few cm from the target, and will be sensitive to the scalar $S$ in the relevant mass range. 
Translating the projected sensitivity to the dark photon parameter space to the case of the leptonic scalar, we arrive at the sensitivity 
reach of HPS shown in Fig.~\ref{Cons}. Above the muon threshold, the scalar decays are too prompt to be detected in this fashion. 
At the same time, muon fixed target experiments have a chance of probing this parameter space for the model. 
This possibility was discussed in Ref.~\cite{Essig:2010gu}, in connection with a possible search for an axion-like particle
in $\mu + {\rm Nucleus} -> \mu + {\rm Nucleus} +a(\to \mu^+\mu^-)$ at the COMPASS facility at CERN \cite{Abbon:2007pq}. Recasting the 
projected sensitivity in the case of the scalar particles, we obtain an $O(1)$ sensitivity to $\xi_l^S$, shown in Fig.~\ref{Cons}.

It is also possible that proton beam dump and fixed target experiments could be sensitive to $S$. Indeed, primary mesons produced subsequently lead to muons, which in turn can radiate the scalar using a larger coupling, $g_\mu$. 
The challenge in such a set-up would be to identify a clean way of detecting electron-positron pairs (or for the case of the fixed target experiments, possibly 
muon pairs) that result from scalar decays. A planned high-energy proton beam dump experiment, SHiP~\cite{Alekhin:2015byh}, as well as 
the existing Fermilab experiment SeaQuest \cite{Gardner:2015wea}, may present advantageous venues, as the high-energy 
and relatively short distance to the detector will increase chances for detecting displaced decays. 

As a separate note, it is worth mentioning that recent studies of the LHCb sensitivity to dark photons \cite{Ilten:2015hya}
may open a new pathway to probe dark scalars as well. The search suggested in \cite{Ilten:2015hya} will not directly apply to a 
leptophilic scalar $S$. Nonetheless, LHCb provides an attractive opportunity to search for  $S$ via its production in association 
with muons. The large boosts available at LHCb may facilitate such searches via displaced decays of $S$.

\subsection{ Future sensitivity from muon decay} 

Flavor-violating muon decays will be scrutinized in a series of upcoming experiments. 
Of particular interest for the model discussed in this paper is the $\mu^+ \to e^+ e^+e^-$ 
search, planned at the Paul Scherrer Institute~\cite{Blondel:2013ia}, which will have exquisite energy resolution 
 for the final state leptons. 

In the present model, the flavor-violating decays of  muons are absent, but the exotic scalars $S$ can be radiated on-shell in the process 
$\mu^+ \to \nu\bar \nu e^+ S \to \nu\bar \nu e^+ e^+ e^-$. The momenta for the electron and one of the two positrons in the  final state must reconstruct the mass of the scalar, $(p_{e^+} + p_{e^-})^2 = m_S^2$. Therefore, a scalar signal would be a 
bump in the invariant mass of the electron-positron pairs, superimposed on the SM background 
$\mu^+ \to \nu\bar \nu e^+ e^+ e^-$. Making use of the recent study of a future 
dark photon search in this set-up \cite{Echenard:2014lma}, we recast the projected 
sensitivity for the case of the leptonic scalar $S$. The signals for $S$ and $V$ were simulated using {\tt MadGraph}. 
For the scalar, emission from the initial muon line dominates, 
since $g_\mu \gg g_e$. The resulting sensitivity reach is shown in Fig.~\ref{Cons}.  

Note that the projections of Ref.~\cite{Echenard:2014lma} assume a prompt decay of the intermediate $e^+ e^-$ resonance. However, for a small portion of the low mass, small $\xi^{S}_\ell$ parameter space where the experiment has sensitivity, the decay length of the $S$ particle can be longer than ${\cal O}$(cm), which is approximately the radius of the innermost silicon detector. Thus, a more careful study must be carried out to assess the sensitivity in this region. The displaced decays may in fact help to reduce the level of background if, of course, the vertex can be cleanly reconstructed. See also Ref.~\cite{Echenard:2014lma} for further discussion of a potential search involving displaced decays.

\subsection{ Kaon decays}

Another well-studied source of muons is via kaon decays. A new particle coupled to muons can be emitted in the decay
$K ^+ \to \mu^+ \nu S$. Note that charge conjugated processes are understood to be implicitly included throughout this section. (For recent discussions of scalar and vector emission in similar processes, 
see Refs.~\cite{Barger:2011mt,Carlson:2013mya}.) For this study, we will concentrate on the past experiment NA48/2~\cite{Fanti:2007vi} and the 
on-going experiment NA62~\cite{Martellotti:2015kna}. 

Depending on the mass of the scalar, it will decay to either $\mu^+\mu^-$ or 
$e^+e^-$. The first case is relatively straightforward.
The SM rate for a similar process, $K ^+ \to \mu^+ \nu \mu^+\mu^-$,
was beyond the reach of previous experiments, and only upper limits on the
corresponding branching fraction exist.
On the other hand, for the electron-positron decays of $S$ there are significant sources of known background. The first source is 
due to a rare SM decay $K ^+ \to \mu^+ \nu e^+e^-$. This process has been measured for 
the invariant mass of a pair in excess of 150 MeV~\cite{Poblaguev:2002ug} with a branching ratio of $7\times 10^{-8}$. 
Below 150 MeV, there is a significant background due to the SM process $K ^+ \to \mu^+ \nu \pi^0$,
with subsequent Dalitz decay of the neutral pion $\pi^0 \to e^+e^-\gamma$ that would mimic the signal if the photon is not detected. 
Finally, there is also some background from pion/muon mis-identification in the underlying $K^+ \to \pi^+ \pi^0$ 
decay and the Dalitz decay of $\pi^0$. 

Even though NA48/2 data has been collected, the corresponding analysis has not yet been done, and
therefore both experiments need to be viewed in terms of potential future sensitivity levels. 
We derive them using the calculated signal rate in our model, and the published detector resolution for 
electron-positron pairs. To estimate the backgrounds, we use known kaon branching ratios and assume that 
the probability of missing a photon is $\sim 10^{-3}$. We also extend $K ^+ \to \mu^+ \nu e^+e^-$ 
to the entire range of $m_{ee}$ using simulations. Above muon threshold we set the rate of the 
signal to 5 events to derive the corresponding sensitivity limits. The projected sensitivity is shown in Fig. \ref{Cons}.

\subsection{ Associated production of scalars with $\tau\bar\tau$ at lepton colliders} 

High-luminosity $B$-factories, such as BaBar and Belle, have collected an integrated luminosity of $\sim 1 ~{\rm ab}^{-1}$, and 
among other things have produced a significant sample of $\tau^+\tau^-$ pairs. The upcoming experiment Belle II 
is aiming to expand this dataset by a factor of ${\cal O}(100)$. Given lepton couplings proportional to mass, 
the associated production of scalars $S$ from the taus,
\be
e^+ e^- \to \tau^+\tau^- + \left(S \to e^+e^- ~{\rm or }~ \mu^+\mu^-\right),
\label{tauprod} 
\ee
 may represent the best chance for discovering or limiting the parameter space for
such particles. The search for exotic particles in association with taus is a relatively unexplored subject, with only one specific 
case analyzed to date~\cite{McKeen:2011aa,BaBar:2014jfk}. 

\begin{figure}
\centerline{\includegraphics[width=0.4\textwidth]{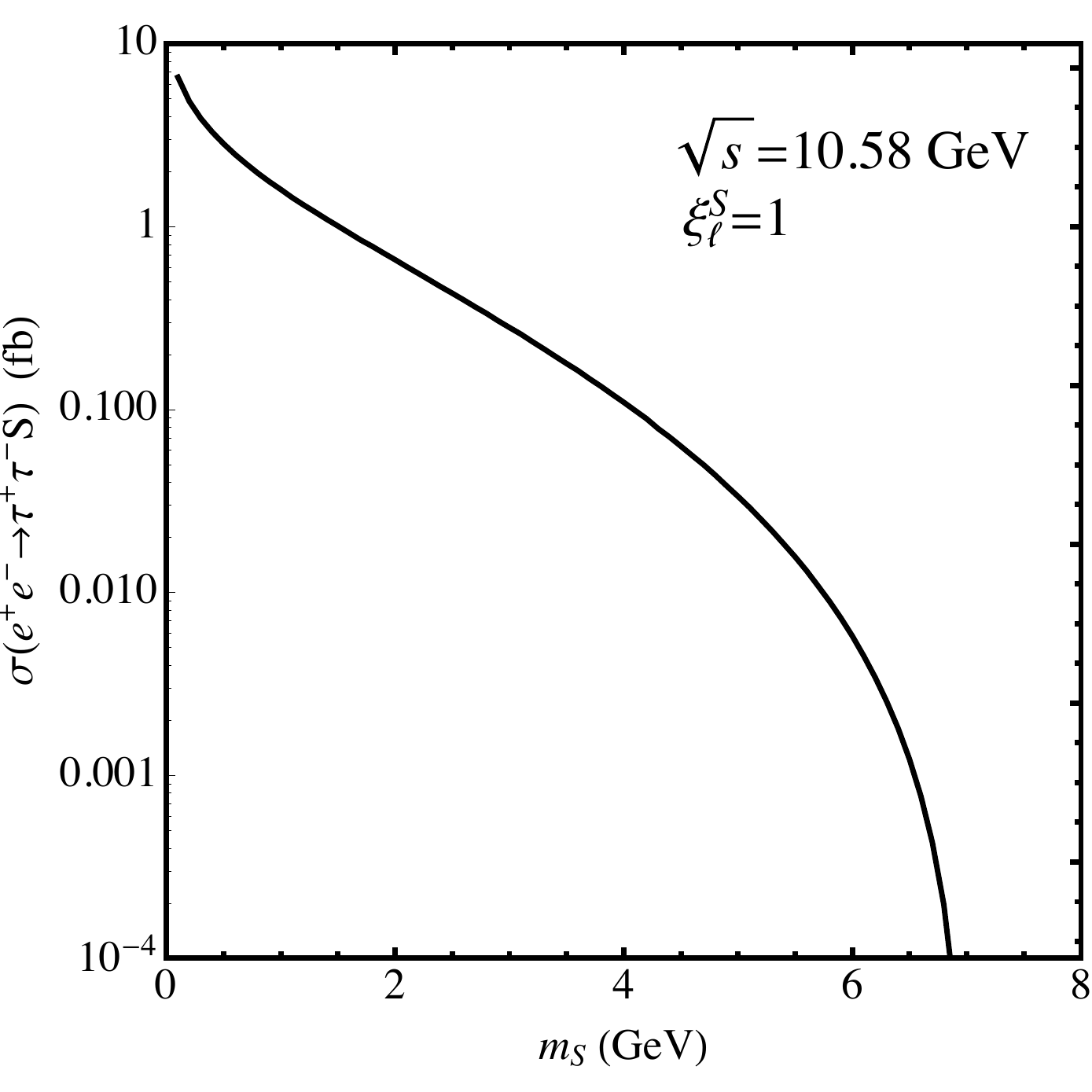}}
\caption{Production rate for $S$ in association with taus at $B$ factories, as a function of $m_S$. The cross section is proportional to $(\xi^S_\ell)^2$, and we have set $\xi^S_\ell=1$.
}
\label{Prod_Bf}
\end{figure}

The production cross section for (\ref{tauprod}) can be calculated analytically. We present the corresponding result as a function of the scalar mass in Fig.~\ref{Prod_Bf}.
To set the scale of the expected event rate for a 1 GeV mass scalar, we take parameters within the muon $g-2$ band, and translate to the 
scale of the coupling to $\tau$-leptons, $g_\tau^2 \sim 1.3\times 10^{-3}$. This leads to a very large number of produced 
scalars in the combined BaBar and Belle dataset, on the order of $5\times 10^4$. Simulating the QED backgrounds using {\tt MadGraph}, 
and requiring that at least one of the taus decay leptonically, we arrive at the sensitivity curves shown in Fig.~\ref{Cons}. 
These sensitivity projections rely on a ``bump hunt" in $\mu^+\mu^-$ (or $e^+e^-$) over the smoothly distributed QED background. 
Notice that for $m_S > 2m_\tau$ the dominant decay mode of the scalar is the tau pair, and the sensitivity is reduced due to the lack of
stable leptons reconstructing to the invariant mass $m_S$. The decay to muons in this mass range is suppressed by $(m_\mu/m_\tau)^2$. 
Also, for scalar masses below $2m_\mu$ the decay length of scalars become comparable to the size of the detector, leading to reduced 
sensitivity. We account for this by introducing a requirement that the $S$ decays occur within $25~{\rm cm}$ of the beam pipe. 

It is worth emphasizing that an analysis of process (\ref{tauprod}) represents perhaps the most effective way of probing the parameter space of the 
leptonic scalar model in a wide mass range, from a few MeV to $\sim 3.5$ GeV.

\section{Constraints on light scalars due to their electroweak properties} 
\label{sec:model_dep}
In this Section we analyze constraints that depend on the embedding of the simple framework of Eq.~(\ref{noUV}) into the SM. We focus on those that are a consequence of our choice of the L2HDM+$\varphi$ scenario outlined in Sec.~\ref{sec:L2HDMS}; in other models, constraints could differ.

\subsection{Higgs decays}

The SM-like Higgs $h$ can decay to pairs of light scalars through both $V_{\rm 2HDM}$ and $V_{\rm portal}$ after electroweak symmetry breaking via the operator $C_{hSS}hSS$.  In the SM-like limit,
\begin{align}
C_{hSS}\simeq\left(\frac{m_h^2}{2\tan\beta}+2m_{12}^2\right)\frac{\left(\xi^S_\ell\right)^2}{v \tan\beta}.
\end{align}
The decays $h\to SS\to4\tau$~\cite{Khachatryan:2015nba,*CMS-PAS-HIG-14-022} and $h\to SS\to2\mu2\tau$~\cite{Aad:2015oqa} have been probed at the LHC, but not observed. These null results can be interpreted as an upper limit on $\xi^S_\ell$. As suggested in~\cite{Curtin:2013fra}, the $2\mu2\tau$ final state offers better reach. In Fig.~\ref{Cons} (right panel), we show the limit from this search for $\tan\beta=200$, $m_{12}=1~\rm TeV$. The constraints become important for the 
muon $g-2$-motivated parameter space once $m_S$ is in the multi-GeV regime.

\subsection{$B$-meson decays}

Although its coupling to quarks and $W$ bosons is suppressed, the scalar mediates quark flavor-changing transitions at one-loop, leading to, for instance, rare $B$ decays like $B\to K\mu^+\mu^-$ (or more generically, $B\to X_s\mu^+\mu^-$) and $B_s\to\mu^+\mu^-$. At large $\tan\beta$ and $\xi^h_\ell=1$, the leading term in the effective Lagrangian mediating $b\to s$ transitions relevant for these decays is
\begin{align}
{\cal L}_{b\to s}\simeq -\frac{3V_{ts}^\ast V_{tb}}{16\pi^2}\frac{m_b m_t^2}{v^3}\frac{m_H^2\xi^S_\ell}{m_h^2\tan^2\beta} S\bar s_R b_L+{\rm h.c.}
\end{align}
This operator can mediate the decay $B_s\to S^\ast\to \mu^+\mu^-$ through an off-shell $S$ and can lead to the decay $B\to KS^{(\ast)}\to K\mu^+\mu^-,Ke^+e^-$. If $2m_e<m_S<2m_\tau$, the decays $B\to KS$ and $B\to K^*S$ can proceed with $S$ decaying to $\mu^+\mu^-$ subsequently; this is subject to strong constraints from the lack of a bump in the $\mu^+\mu^-$ invariant mass in $B\to K^\ast\mu^+\mu^-$ at LHCb~\cite{Aaij:2015tna}. We show limits on $\xi^S_\ell$ that result from these decay modes in Fig.~\ref{Cons}, taking $\tan\beta=200$, $m_{H}=m_{H^\pm}=500~\rm GeV$. (The degeneracy of the heavy Higgs masses weakens electroweak precision constraints.) Notice that for the mass range $2 m_\tau < m_S < m_B - m_K^{(*)}$ the sensitivity is degraded as 
$S$ would primarily decay to a tau pair. 

We note in passing that the constraint on $\xi^S_\ell$ could be weakened by a factor $\sim m_H^2/m_h^2$ if $\xi^h_\ell\sim-1$ [cf. Eq.~(\ref{eq:deltas})] which is consistent with the data on Higgs properties.

For $m_S< 2 m_\mu$ the important search channels are $B\to X_se^+e^-$. These modes are better suited for searches at  Belle II, 
and sensitivities below branchings of $10^{-8}$ will also cover the remaining `triangular' parameter space in Fig.~\ref{Cons} (right panel). 

\subsection{Electroweak precision constraints} 

Enhanced couplings of the lepton-specific Higgses will also induce one-loop corrections to leptonic branching ratios of the $Z$-boson. Here we analyze 
$R_\tau$, defined as  $R_\tau \equiv {\Ga(Z\rightarrow {\rm hadrons})}/{\Ga(Z\rightarrow \tau\bar{\tau})}$,
where $\Ga(Z\rightarrow {\rm hadrons}) \propto \sum_{q=u,d,s,c,b} (|g_{q_L}|^2+|g_{q_R}|^2)$ and $\Ga(Z\rightarrow \ta\bar{\ta}) \propto (|g_{\ta_L}|^2 + |g_{\ta_R}|^2)$,
with $g_L = I_3 -Q s^2_W$ and $g_R=-Q s^2_W$. $s_W$ stands for the sine of the weak mixing angle.
Perturbations to  $R_\ta$ can be expressed in terms of corrections to $s^2_W $ and modifications of the $Z\tau\tau$ vertices by the scalar loops, 
\ba
 \frac{\De R_\ta}{R_\ta}  &=& 4.3 \de g_{\ta_L} - 3.7 \de g_{\ta_R} - 0.8 \de s^2_W \\
  &\simeq& 4\de g_A^\ta + 1.9\times 10^{-3} T \label{corr}
\ea
with $g_A = g_L - g_R$.

Interpreting the PDG fit, $ R_\tau = 20.764 \pm 0.045 $ as the constraint, $-2\times 10^{-3} \leq \De R_\ta/R_\ta \leq 2 \times 10^{-3}$, we compare it to the result of the one-loop calculation in our model. The corrections to $\de g_{\ta_L}$ and $\de g_{\ta_R}$ can be obtained in the L2HDM model following \cite{LLT1,LLT2}, and we present the ensuing constraint in the right panel of Fig.~\ref{Cons}. The contributions due to loops of scalars that are (mostly) components of electroweak doublets are negligible for $m_{H,H^\pm,A}\gtrsim300~\rm GeV$, even for $\tan\beta$ as large as 200, as taken in Fig.~\ref{Cons}.

Additionally, we mention that as long as there is some degeneracy in the masses of at least two heavy scalars (at the order of $\sim 50~\rm GeV$), corrections to the oblique electroweak parameters $S$, $T$, and $U$ are not constraining.

\section{Discussion and conclusions} 
\label{sec:concl}
We have analyzed a simplified model of a light `dark scalar' that couples predominantly to 
leptons. This hidden sector model has a very distinct phenomenology, differing in several ways from the 
phenomenology of the canonical dark photon model. It is interesting that the coupling of a light scalar $S$ to leptons can still be of order $m_\mu/v$, 
and thus capable of inducing a large shift in the anomalous magnetic moment of the muon, without being excluded by direct searches. 
This is because the coupling to electrons relative to muons is 
suppressed by $m_e/m_\mu$, and many constraints that have ruled out the minimal version of the 
dark photon model as an explanation of the muon $g-2$ discrepancy do not have any constraining power. 

The simplified model (\ref{noUV}) does not, however, respect the SU(2)$\times$ U(1) gauge symmetry of the 
SM and needs a UV completion. This implies that either the field $S$ or the fermion fields in (\ref{noUV}) cannot 
have well defined charge assignments. One possible UV completion, investigated in this paper, defines $S$ predominantly 
as a singlet scalar with a small admixture of an SU(2) doublet. On the other hand, one can consider the possibility of 
 lepton fields in (\ref{noUV}) arising from a mixing between the `normal' SM fields and heavy 
vector-like leptons \cite{Chen:2015vqy}, so that mixing with a pure singlet $S$ becomes possible. 

The UV completion of the model proposed here is based on the lepton-specific two Higgs 
doublet model, augmented by an additional light singlet. In the large $\tan \beta$ regime, the Yukawa 
couplings of the lepton-specific Higgs bosons $h_l \supset (H,A, H^\pm)$ 
to leptons are enhanced relative to their SM values. 
If an {\em additional} singlet field $\varphi$ mixes with $h_l$, the end result can be a new 
light boson $S$ with couplings to leptons that scale as $m_l$ and are of order 
 the SM Yukawa couplings, proportional to the product of a small mixing angle $\theta$ and 
large $\tan \beta$. At the same time, the couplings of $S$ to quarks and weak gauge bosons are
suppressed, which softens all constraints from the FCNC processes derived from $K$, $B$ physics. 
Moreover, there are no charged lepton flavor violating processes, since flavour conservation 
is built into the Yukawa structure of the model. (For the alternate UV completion with vector-like fermions
\cite{Chen:2015vqy}, flavor symmetry in the charged lepton sector
is likely to be broken. At the same time, the pure singlet nature of $S$ in this type of UV completion may allow flavor changing processes to 
be kept separate for the quark and lepton sectors, thus avoiding strong constraints from hadronic FCNC.)

We have analyzed a wide selection of constraints and sensitivity limits from the existing experiments, and 
from upcoming searches. The production of scalars is enhanced in processes that involve muons and 
tau leptons. We have studied muon and kaon decays, and shown that future experiments and analyses of 
the existing data ({\em e.g.} by NA48/2, BaBar and Belle experiments) are capable of reaching the levels of sensitivity 
to the parameter space suggested by the muon $g-2$ discrepancy. The mass range $m_S <2 m_\mu$ 
naturally leads to longer lived bosons, and may be probed through experiments that have sensitivity to 
displaced decays, such as the HPS experiment at JLab. 

Perhaps the most sensitive current search for a leptonic dark scalar can be performed by the BaBar and Belle collaborations, using existing data. 
The process of interest involves tau pair production with an associated emission of the scalar. 
The large datasets generated by the two experiments will allow a sensitive analysis of $\tau^+\tau^-\mu^+\mu^-$
and $\tau^+\tau^-e^+e^-$ production, looking for a peak in the invariant mass of electrons and muons. 
Even without extra data that should be collected at Belle~II, the two $B$-factories should comprehensively test the 
dark scalar model in the wide mass range spanning  almost three orders of magnitude. 

The constraints and projected sensitivity reach for many experiments are summarized in the two panels of Fig.~\ref{Cons}. The
results in the left panel are based only on the simplified model (\ref{noUV}) and use only the 
$g_l \propto m_l$ scaling and absence of invisible decay channels for $S$. Much stronger constraints 
are derived for $m_S>2 m_\mu$ using quark flavor physics, within the lepton-specific 2HDM UV completion. 
One should still keep in mind that the strong constraints shown in the right panel of Fig. \ref{Cons} are indeed 
very sensitive to the type of UV completion, and can in principle be avoided with a different microscopic model of (\ref{noUV}). 

{\em Note added:} Following the completion of this work, the BaBar Collaboration released a preprint~\cite{BaBar} with an analysis that constrains any light vector particle ($V$) in the $e^+e^-\to \mu^+\mu^-V \to \mu^+\mu^-\mu^+\mu^-$ channel. This limit can be appropriately recast for the scalar model, and we show the resulting constraint as the solid black line in Fig. 1. This is now the strongest model-independent constraint over a large region of the $2 m_\mu < m_S < 2 m_\tau$ mass range. However, unlike the case of a vector coupled to $L_\mu-L_\tau$, the limit from~\cite{BaBar} does not rule out the $g-2$ band in that region. The reason is that the scalar contribution to $g-2$ is somewhat larger than that of the vector and its production cross section is smaller at the same mass and coupling to muons. The constraint can be improved even further if BaBar performs the corresponding $e^+e^-\to \tau^+\tau^-S$ analysis. 

\acknowledgements

We would like to thank B. Echenard, E. Goudzovski, I. Nugent, M. Roney and B. Shuve for helpful discussions. The work of  M.\,P. and A.\,R. is supported in part by NSERC, Canada, and research at the Perimeter Institute is supported in part by the Government of Canada through NSERC and by the Province of Ontario through MEDT. The work of B.\,B. is supported in part by the U.S. Department of Energy under grant No. DE-SC0015634. The work of D.\,M. is supported by the U.S. Department of Energy under grant No. DE-FG02-96ER40956.


\bibliography{LHport}


\end{document}